\documentclass[11pt]{article}
\usepackage{amsmath,color}
\usepackage{amssymb}
\usepackage{amsfonts}
\usepackage{graphicx}
\usepackage{color}

\definecolor{darkgreen}{rgb}{0,0.35,0}

\numberwithin{equation}{section}

\begin{document}

\title{{$T\bar{T}$-deformations, AdS/CFT and correlation functions}}
\author{Gaston Giribet}
\maketitle

\begin{center}

\smallskip
\smallskip
\centerline{Center for Cosmology and Particle Physics, New York University}
\centerline{{\it 726 Broadway 10003 New York City, USA}}

\end{center}

\bigskip

\bigskip

\bigskip

\bigskip

\bigskip

\begin{abstract}
A solvable irrelevant deformation of AdS$_3$/CFT$_2$ correspondence leading to a theory with Hagedorn spectrum at high energy has been recently proposed. It consists of a single trace deformation of the boundary theory, which is inspired by the recent work on solvable $T\bar{T}$ deformations of two-dimensional CFTs. Thought of as a worldsheet $\sigma $-model, the interpretation of the deformed theory from the bulk viewpoint is that of string theory on a background that interpolates between AdS$_3$ in the IR and a linear dilaton vacuum of little string theory in the UV. The insertion of the operator that realizes the deformation in the correlation functions produces a logarithmic divergence, leading to the renormalization of the primary operators, which thus acquire an anomalous dimension. We compute this anomalous dimension explicitly, and this provides us with a direct way of determining the spectrum of the theory. We discuss this and other features of the correlation functions in presence of the deformation.
\end{abstract}

\newpage

\section{Introduction}

In \cite{GIK1}, the question of constructing a holographic dual to asymptotically linear dilaton backgrounds was reconsidered in the light of recent studies of solvable irrelevant deformations of two-dimensional conformal field theory (CFT) \cite{SmirnovZamolodchikov, Cavaglia}. This relates to the vacua of the so-called little string theory (LST) \cite{Berkooz}, a non-local, non-gravitational theory which exhibits Hagedorn spectrum at high energy \cite{Pelc, Hagedorn}. From the bulk point of view, this theory corresponds to a locally flat background configuration with a dilaton that grows linearly in the weak coupling region. The analysis carried out in \cite{GIK1} addresses the question as to how to realize this type of theory from the holographic point of view, aiming at achieving in such a way a microscopic description of the Hagedorn spectrum. The proposal is that the holographic dual to string theory in asymptotically linear dilaton background is given by a particular single-trace deformation of the symmetric product CFT that appears in the holographic description of string theory on AdS$_3 \times S^3 \times T^4$ space. This deformation is inspired by the solvable $T\bar{T}$-deformation studied in \cite{SmirnovZamolodchikov, Cavaglia} (see also \cite{Sergei1, Sergei2}) and shares many features with it, such as universality and solvability. However, unlike the deformations considered in \cite{SmirnovZamolodchikov, Cavaglia}, the one in \cite{GIK1} rather consists of a single-trace deformation and, as such, induces a local deformation of the bulk geometry. In fact, the interpretation of the result of \cite{GIK1} is that it gives a dual theory that interpolates between a standard CFT$_2$ in the infrared (IR), which is dual to the AdS$_3$ strings, and a non-local theory with Hagedorn spectrum in the ultraviolet (UV), which would be dual to the linear dilaton background. In this sense, this may be regarded as a solvable irrelevant deformation of AdS$_3$/CFT$_2$ correspondence \cite{GIK2}.

The starting point of the setup considered in \cite{GIK1} is superstring theory on AdS$_3\times S^3\times T^4$ spacetime with NS-NS fluxes, the background that describes the near-horizon region of the NS1-NS5 system. The corresponding string $\sigma$-model admits an exact description in terms of the Wess-Zumino-Witten (WZW) model formulated on $SL(2,\mathbb{R})\times SU(2) \times U(1)^4$, where the WZW level, $k$, is given by the number of five-branes that wrap on the $T^4\times S^1$ subspace. There are also fundamental strings, say $p$ of them, which are wrapped on the isolated $S^1$. In the large $p$ limit, string theory on AdS$_3\times S^3$ becomes weakly coupled. Large $k$ regime corresponds to the limit in which the string length scale $\sqrt{\alpha '}$ is small in comparison with the radius of the AdS$_3$ (and the $S^3$) space(s). This type of AdS$_3$ solution to string theory has been extensively studied in the literature \cite{GKS, Seiberg, AdS, MO1, GN3, MO3} and it represents one of the few examples in which holography can be explored beyond the supergravity approximation, allowing to have access to purely stringy effects. In fact, precision tests of AdS/CFT at finite $\alpha '$ (i.e. finite $k$) have been carried out for this background \cite{Gaberdiel, Pakman, Rastelli, Cardona}. String theory on AdS$_3\times S^3\times T^4$ is closely related to two-dimensional vacua of LST, in a way that is nicely explained in \cite{Pelc} and references thereof; see also \cite{GIK1, GIK2}. This is why this is the natural setup to consider.

In this paper, we consider the setup of references \cite{GIK1, GIK2}. For this type of solvable irrelevant deformation of AdS$_3$/CFT$_2$, which we review in sections 2 and 3, we compute the worldsheet 2-point correlation functions of Virasoro primary operators. This provides a direct way of determining the spectrum of the deformed CFT by reading the anomalous dimension of the primary states. We compute these 2-point functions in two different basis of vertex operators, and resorting both to path integral techniques and to perturbation theory. We do this in sections 4 and 5, respectively; and in section 6 we comment on the pole structure of the 2-point function. In section 7, we compare the spectrum obtained from the correlation functions with the coset description of it proposed in \cite{GIK2}, showing the agreement. In section 8, we comment on the 3-point function and how the same techniques lead to write higher-point functions of the deformed theory in terms of correlation functions of Liouville field theory. In section 9, we briefly discuss how the correlators can also be efficiently computed by thinking of the worldsheet theory as a $\sigma$-model on $(SL(2,\mathbb{R})/U(1))\times U(1)$. Section 10 contains some final remarks.

\section{$T\bar{T}$ type deformation}

The deformation proposed in \cite{GIK1}, which interpolates between the IR CFT$_2$ and the UV non-local theory, is given by an irrelevant operator built out of the holomorphic and anti-holomorphic components of the boundary CFT$_2$ stress tensor written down in \cite{Seiberg}. The holomorphic (resp. anti-holomorphic) component of such stress tensor is given by a second order differential operator of the $SL(2,\mathbb{R})$ isospin variable acting on a bilinear arrange of the $SL(2,\mathbb{R})\times SL(2,\mathbb{R})$ currents of the WZW model multiplied by a normalizable operator of weight $1$, $\Phi_{h=1}$ (see \cite{Seiberg, GIK1} for details). After integrating over the isospin variable, $x$, and treating the boundary terms adequately, the deformation operator takes the form
\begin{equation}
D= \frac{\lambda_0}{\pi }\int d^2z J^-\bar{J}^- \label{D}
\end{equation}
where $J^-$ and $\bar{J}^-$ are the local currents corresponding to the lower triangular generators of the $\hat{sl}(2)_k\oplus \hat{sl}(2)_k$ affine algebra. Variables $z$, $\bar{z}$ are worldsheet variables. In order to understand what operator (\ref{D}) means from the dual theory point of view, recall that the zero mode of the local current $J^-$ (and $\bar{J}^-$) corresponds in the boundary IR CFT to the $SL(2,\mathbb{R})$ generator $L_{-1}$ (resp. $\bar{L}_{-1}$). In turn, while (\ref{D}) represents a marginal operator in the worldsheet CFT$_2$, as demanded by consistency of string theory\footnote{The marginal deformation (\ref{D}) has been previously studied in \cite{Israel}.}, it represents an irrelevant dimension-$(2,2)$ deformation of the boundary CFT$_2$. The coupling constant $\lambda _0$ of this deformation emerges from the integration of the normalizable operator $\Phi_{h=1}$ over the isospin variable $x$, variable that is usually introduced to label the $SL(2,\mathbb{R})$ representations and is connected to the usual eigenvalue $m$ of the $SL(2,\mathbb{R})$ Cartan generator $J^3$ by a Mellin transform \cite{Seiberg}. More precisely, if one organizes the $SL(2,\mathbb{R})$ local currents $J^{3,\pm }$ (and $\bar{J}^{3,\pm }$) in the polynomial form $J=2xJ^{3}-J^{+}-x^2J^{-}$ (resp. $\bar{J}=2\bar{x}\bar{J}^{3}-\bar{J}^{+}-\bar{x}^2\bar{J}^{-}$), the holomorphic and anti-holomorphic components of the stress-tensor of the boundary CFT$_2$ can be written as \cite{Seiberg}
\begin{equation}
T(x)=\frac{\pi}{2k} \int d^2z (\partial_x J\partial_x+2\partial_x^2J) \Phi_{h=1} \bar{J},
\ \ \ \bar{T}(\bar{x})=\frac{\pi}{2k}\int d^2z (\partial_{\bar{x}} \bar{J} \partial_{\bar{x}}+2\partial_{\bar{x}}^2\bar{J})\Phi_{h=1}{J}, \label{Dtruco} 
\end{equation}
where each component turns out to be given by an integral over the worldsheet. $\Phi_{h=1}$ is a normalizable dimension-zero operator in the worldsheet, which corresponds in the boundary to an operator of dimension $h=1$. Complex variables $x,\bar{x}$ can be thought of as the coordinates of the boundary where the dual CFT$_2$ lives. The deformation (\ref{D}) rather corresponds to a single-trace version of (\ref{Dtruco}), in the sense that it consists in a similar operator but with the holomorphic and anti-holomorphic pieces integrated together over the worldsheet variable; namely
\begin{equation}
D = \frac{\pi}{2k} \int d^2x \int d^2z \ (\partial_x J\partial_x+2\partial_x^2J) (\partial_{\bar{x}} \bar{J} \partial_{\bar{x}}+2\partial_{\bar{x}}^2\bar{J})\Phi_{h=1}. 
\end{equation}
After integration in $x$, this yields the local operator in (\ref{D}) with the coupling constant $\lambda_0\sim \int d^2x\ \Phi_{h=1}$; see \cite{GIK1}. As we will see, $\lambda_0$ appears in the anomalous dimension of the vertex operators. It has a clear geometrical interpretation from the bulk point of view as it sets the scale where the geometric transition between AdS$_3$ and the linear dilaton background takes place\footnote{The specific scale $\lambda_0$ is actually meaningless as its value can be changed using symmetry (\ref{Symetron}) below. What is meaningful are the behaviors the theory exhibits in the two limits $\phi \to \pm \infty$}.

The index $h$, which labels the $SL(2,\mathbb{R})$ representations, corresponds in the boundary to the scaling dimension of the dual operators, $\hat{\mathcal{O}}_h$, of the IR CFT. We are interested in computing the 2-point correlation function of two such operators, $\langle \hat{\mathcal{O}}_h(p)\hat{\mathcal{O}}_h(-p)\rangle$, in the boundary theory. However, given the correspondence \cite{Gaberdiel, Pakman, Rastelli, Cardona} with the 2-point correlation numbers in the bulk, $\langle \hat{\mathcal{O}}_h(p)\hat{\mathcal{O}}_h(-p)\rangle =\int d^2z_1 d^2z_2 \text{Vol}^{-1}_{PSL(2,\mathbb{C})}\langle \Phi_{h}(p|z_1)\Phi_{h}(-p|z_2)\rangle$, we will instead perform the computation on the worldsheet theory, taking profit form the fact that, remarkably, the worldsheet theory in presence of the deformation is still exactly solvable. The basis of primary operators $\Phi_h(p|z)$ we will consider first is the one proposed in \cite{GIK2}, which satisfy the operator product expansion (OPE)
\begin{equation}
J^-(z)\Phi_h(p|z_i)\simeq \frac{ip}{(z-z_i)} \Phi_h(p|z_i) + ...
\end{equation}
where the ellipses stand for regular terms. Since these operators create eigenstates of the currents $J^-$, $\bar{J}^-$, one could naively expect the correlation function of two such operators to satisfy the equation
\begin{equation}
\frac{d}{d\lambda_0}\langle \Phi_h(p|z_1) \Phi_h(-p|z_2) \rangle \sim - \frac{|p|^2}{\pi } \langle \Phi_h(p|z_1) \Phi_h(-p|z_2) \rangle
\end{equation}
and thus to behave like
\begin{equation}
\langle \Phi_h(p|z_1) \Phi_h(-p|z_2) \rangle \sim \frac{e^{-\frac{\lambda |p|^2}{\pi}}}{|z_1-z_2|^{4\Delta }}, \label{2pf}
\end{equation}
with $\Delta $ being the worldsheet conformal dimension, which {a priori} depends on $h$ and $p$. $\lambda $ is the coupling $\lambda_0$ dressed by a factor. However, this argument is unjustifiably fast, as the insertion of the operator $D$ in the 2-point function produces a logarithmic singularity and leads to the renormalization of the vertex operators, which acquire an anomalous dimension. In other words, the dressing factor of the coupling constant (namely the ratio $\lambda /\lambda_0 $) results infinite due to a logarithmically divergent integral (see (\ref{SV0}) below). Therefore, in order to derive an expression for the 2-point function one has to be more cautious and undertake the computation of the expectation values explicitly. We will do this below in two different basis and by two different methods. From the expressions obtained, we will read the anomalous dimension of the operators. This will provide us with a direct method to compute the spectrum of the theory. We will compare the result with the coset construction of \cite{GIK2} showing the consistency. Finally, we will explain how the method used here to compute the 2-point correlation function can be applied to compute higher-point correlation functions in the deformed CFT.

\section{The worldsheet theory and its deformation}

To represent the AdS$_3$ sector, whose worldsheet theory is described by the $SL(2,\mathbb{R})$ WZW model, we consider, as in \cite{GIK1, GIK2}, the Wakimoto representation \cite{Wakimoto}. This amounts to write the WZW action, perturbed by the operator $D$, as follows 
\begin{equation}
S = \frac{1}{2\pi }\int d^2z \Big( \partial \phi \bar{\partial } \phi - \sqrt{{2}/{k}}{\mathcal R}\phi -\beta \bar{\partial }\gamma - \bar{\beta }\partial \bar{\gamma }-2M_0 \beta\bar{\beta } e^{-\sqrt{{2}/{k}}\phi  } - 2\lambda_0 \beta\bar{\beta } \Big) \ , \label{S}
\end{equation}
which includes quantum corrections. This action describes a perturbation of a free CFT with a non-compact boson $\phi\in \mathbb{R}_{\geq 0}$ with background charge $\sqrt{k/2}$, and a (1,0) $\beta $-$\gamma$ commutative ghost system. The non-trivial correlators in the free CFT are $\langle \phi(z) \phi(0) \rangle = -2\log |z| $, $\langle \beta(z) \gamma(0) \rangle = 1/z$. For computational reasons, it might be convenient to bosonize the $\beta $-$ \gamma$ system by defining $\beta= -i\partial v e^{iv-u} , $ $\gamma = e^{u-iv}$ with $\langle u(z) u(0) \rangle = \langle v(z) v(0) \rangle = -2\log |z|$. In terms of the fields $u$, $v$ the combinatorial game when Wick contracting fields inside the correlation functions gets notably simplified. 

The undeformed AdS$_3$ $\sigma $-model corresponds to $\lambda_0 = 0 $ in (\ref{S}). In these variables, the metric of the Poincar\'{e} patch of AdS$_3$ takes the form 
\begin{equation}
ds^2=R^2 (d\phi^2+e^{2\phi }d\gamma d\bar{\gamma}) =\frac{R^2}{r^2}({dr^2+dX^2-dT^2})
\end{equation}
where $R^2=k\alpha'$, $r=e^{-\phi }$, $\gamma = X-T$, $\bar\gamma = X+T$. The near boundary region of the space corresponds to large $\phi$, i.e. $r\sim 0$.

The last term in (\ref{S}) is the deformation (\ref{D}), as in the Wakimoto representation the local current $J^-$ is simply given by the field $\beta$, and analogously for the anti-holomorphic sector. $\beta$ and $\bar{\beta}$ are auxiliary fields, and so they can be integrated out. If doing so, the action takes the form
\begin{equation}
S = \frac{1}{2\pi }\int d^2z \Big( \partial \phi \bar{\partial } \phi - \sqrt{{2}/{k}}{\mathcal R}\phi 
+\frac{e^{\sqrt{{2}/{k}}\phi} }{2+2\lambda_0 e^{\sqrt{2/k}\phi} } {\partial\bar{\gamma}\bar{\partial}\gamma} \Big), \label{SD}
\end{equation}
where, in addition, we have shifted the zero-mode of the dilaton as $\phi\to\phi+\sqrt{k/2}\log M_0$ in order to absorb the value of $M_0$. To be more precise, the dilaton configuration, while behaving linearly at large $\phi$, also receives corrections for finite $\lambda_0$. This can be seen as follows: In the bosonic theory with $\lambda_0=0$, quantum corrections produce the linear dilaton term after integrating on the $\beta $-$\gamma$ fields \cite{Satoh}. The original Gaussian functional measure of the WZW $\sigma $-model actually corresponds to $\mathcal{D}{\phi }\mathcal{D}(e^{\phi}\gamma)\mathcal{D}(e^{\phi}\bar{\gamma})e^{-S_{\text{WZW}}}$ with $S_{\text{WZW}}={(k/\pi )\int d^2z(\partial \phi \bar{\partial }\phi +2e^{2\phi}\partial \bar{\gamma}\bar{\partial}\gamma)}$ being the the classical WZW action. Then, in the process of transforming this measure into $\mathcal{D}{\phi }\mathcal{D}^2\gamma e^{-S_{\text{eff}}}$ a Jacobian $J=e^{S_{\text{WZW}}-S_{\text{eff}}}$ appears. The logarithm of this Jacobian takes the form $\log J=({2}/{\pi})\int d^2z(\partial \phi \bar{\partial }\phi -(1/2)\mathcal{R}\phi )$. By rescaling $\phi \to \phi /\sqrt{2(k-2)}$ one eventually finds the effective action (\ref{SD}) with a shifted level $k-2$ instead of $k$, with $\lambda_0=0$ and $M_0=2k$. In the supersymmetric theory, the shifting of the level does not occur. In the case $\lambda_0 \neq 0$, the coupling between $\gamma $, $\bar{\gamma }$, and $\phi$ fields is more involved, and the effective action acquires a dilaton configuration that goes linearly only at large $\phi$ (see \cite{GIK1} for the discussion about the corresponding supergravity solution). 

The $\sigma$-model (\ref{SD}) describes a string propagating in a space that interpolates between a linear dilaton background in the region $\lambda_0 \gg e^{-\sqrt{2/k}\phi}$, say in the limit $\phi\to\infty $, and AdS$_3$ background in the opposite limit, $\phi\to-\infty$. That is to say, the UV behavior in the dual theory is governed by the linear dilaton theory, while AdS$_3$ emerges in the IR. In the former case, action (\ref{SD}) takes the form $
S = {1}/({2\pi })\int d^2z ( \partial \phi \bar{\partial } \phi - \sqrt{{2}/{k}}{\mathcal R}\phi 
+ \partial X \bar{\partial }X - \partial T \bar{\partial }T )
$, where we have rescaled coordinates $X$ and $T$ by a factor $\sqrt{2\lambda_0}$. This describes the flat dilatonic background \cite{Berkooz, Pelc}. In the latter case, the model becomes the standard AdS$_3$ $\sigma$-model \cite{GKS, Seiberg, AdS, MO1, GN3, MO3}. 

It is worth noticing that action (\ref{S}) is symmetric under the transformations 
\begin{equation}
\beta \to e^{\sigma }\beta , \ \ \ \gamma \to e^{-\sigma }\gamma , \ \ \ \phi \to \phi +\sqrt{2k}\sigma , \ \ \ \lambda_0\to e^{-2\sigma }\lambda_0, \label{Symetron}
\end{equation}
with $\sigma $ being a real constant. It is also symmetric under 
\begin{equation}
\phi \to \phi +\sqrt{2k}\sigma , \ \ \  M_0\to e^{2\sigma }M_0.
\end{equation}
These symmetries can be used to set the values of the coupling constants by shifting the dilaton zero-mode, namely $\phi_0$. Direction $\phi$, which is semi-infinite, is in fact the direction along which the dilaton grows linearly. The spectrum has the characteristic gap in the momentum along that direction. The momentum (in string units) along $\phi $ is parameterized by $h$, with $\Phi_h(p|z)\sim e^{\sqrt{k/2}(h-1)\phi}$ and
\begin{equation}
h=\frac{1}{2} + \frac{i}{2} {\sqrt{k}}p_{\phi } \label{LAGHY}
\end{equation}
with $p_{\phi}\in \mathbb{R}$. These values oh $h$ correspond to the continuous series of the $SL(2,\mathbb{R})$ representations. 

The bosonic part of the vertex operators in the AdS$_3$ subspace is given by
\begin{equation}
\Phi_h(p|z)= e^{ip\gamma(z)+i\bar{p}\gamma(\bar{z})} e^{\sqrt{{2}/{k}}(h-1)\phi(z)}\label{24}
\end{equation}
This basis is similar to the $\mu $-basis introduced in reference \cite{RT}; see also \cite{HS}. This also appears in the description of string theory on the three-dimensional black hole \cite{Ross}.

\section{Correlation functions: path integral computation}

We consider the correlation functions
\begin{equation}
\Big\langle {\Phi}_{h}(p_1|z_1) {\Phi}_{h}(p_2|z_2) \Big\rangle =  
\int \mathcal{D}^2\beta \mathcal{D}^2\gamma \mathcal{D}\phi \ e^{-S}{\Phi}_{h}(p_1|z_1) {\Phi}_{h}(p_2|z_2) \label{GHJ}
\end{equation}

These functions can be computed using the techniques developed in \cite{HS}. Those techniques allow to reduce the calculation of any correlation functions in the $SL(2,\mathbb{R})$ WZW model to a computation in Liouville theory, which is simpler. The trick goes as follows: Integrating by part the kinetic term of the $\beta $-$ \gamma$ system, using the fact that $ {\Phi}_{h}(p|z)\sim e^{ip\gamma }$, the functional integral over $\gamma $ and $\bar{\gamma }$ can be explicitly performed because each of these fields only appears once in each term of the Lagrangian. This yields a $\delta$-function resulting in the following equations
\begin{eqnarray}
\bar{\partial }\beta(z) &=& 2\pi i \Big(p_1\delta^{(2)} (z-z_1)+p_2\delta^{(2)} (z-z_2)\Big) \ ,
\end{eqnarray}
and analogously for $\partial\bar{\beta}$ with $p_{1,2}\leftrightarrow\bar{p}_{1,2}$. Here, $\delta^{(2)}(z)=\delta(z)\delta(\bar{z})$. Taking into account that $\bar{\partial}(1/z)=2\pi \delta^2(z)$, one finds the following solution for $\beta $
\begin{equation}
\beta (z) = \frac{ip_1}{(z-z_1)} + \frac{ip_2}{(z-z_2)} \label{34}
\end{equation}
and analogously for $\bar{\beta }$. Since $\beta $ and $\bar{\beta }$ are assumed not to have additional singularities at $0\neq z \neq 1$ on the whole Riemann sphere $\mathbb{C}\cup \{\infty \}$, the solution to (\ref{34}) only exists if $p_1+p_2=0$. That is, since $\beta $ is a meromorphic 1-differential on $\mathbb{CP}^1$ the general solution finally takes the form
\begin{equation}
\beta (z)= \frac{ip_1(z_1-z_2)}{(z-z_1)(z-z_2)}. 
\end{equation}

Plugging this back in the action produces the following two terms in the effective potential
\begin{equation}
2 |p_1|^{2} |z_1-z_2|^2\int d^2z \ {|z-z_1|^{-2}|z- z_2|^{-2}} \Big (M_0 \ e^{-\sqrt{{2}/{k}}\phi }+\lambda_0 \Big) .\label{tr}
\end{equation}

Then, by redefining the field $\phi$ as follows
\begin{equation}
\phi (z)\to \phi (z)- \sqrt{2k}\Big( \log|z-z_1|+\log|z-z_2|-\log |p_1|-\log |z_1-z_2|\Big) , \label{shift}
\end{equation}
one can absorb the $z$- and $z_i$-dependence in the first term of (\ref{tr}), and this suffices to turn such term into a Liouville wall operator
\begin{equation}
2M_0  \int d^2z \ e^{\sqrt{{2}}b\phi } \label{Liouville}
\end{equation}
with $b=-1/\sqrt{k}$. (This also shifts the background charge and the momenta of the vertex operators in a certain way \cite{HS}). The issue is with the second term of (\ref{tr}), in which the $z_i$-dependent factors remain in the integrand and actually produce a divergence. In fact, integral
\begin{equation}
I_0=|z_1-z_2|^2 \int_{\mathbb{C}} d^2z \ {|z-z_1|^{-2}|z-z_2|^{-2}} \label{SV0}
\end{equation}
is logarithmically divergent, as power counting obviously indicates. In order to see this explicitly, one can first introduce a regulator $\epsilon$ and write
\begin{equation}
I_{\epsilon }=|z_1-z_2|^2 \int_{\mathbb{C}} d^2z \ {|z-z_1|^{2\epsilon-2}|z-z_2|^{2\epsilon-2}}=|z_1-z_2|^{4\epsilon } \int_{\mathbb{C}} d^2x \ {|x|^{2\epsilon-2}|x-1|^{2\epsilon-2}} \label{qUltima}
\end{equation}
which permits to keep control of the divergence. We have $\lim_{\epsilon\to 0}I_{\epsilon}=I_0$. In the second equality in (\ref{qUltima}), we used $x=(z-z_1)/(z_2-z_1)$. The integral over $x$ is the Shapiro-Virasoro integral 
\begin{equation}
\int_{\mathbb{C}} d^2x \ {|x|^{2a_1-2}|x-1|^{2a_2-2}} = 2 \pi \frac{\Gamma(a_1)\Gamma(a_2)\Gamma(1-a_1-a_2)}{\Gamma(1-a_1)\Gamma(1-a_2)\Gamma(a_1+a_2)}
\end{equation}
for parameters $a_1=a_2=\epsilon$. This formula clearly diverges in the limit $\epsilon\to 0$ since, in that limit $\Gamma(\epsilon) \sim  ({1}/{\epsilon}) + const + \mathcal{O}(\epsilon)$, and then
\begin{equation}
\int_{\mathbb{C}} d^2x \ {|x|^{2\epsilon-2}|x-1|^{2\epsilon-2}} \simeq \frac{4 \pi}{\epsilon} + const.
\end{equation}
up to terms that vanishes at $\epsilon =0$. Therefore, expanding expression (\ref{qUltima}) in powers of $\epsilon $, one finds
\begin{equation}
I_{\epsilon }=\Big( 1 + 4\epsilon \log|z_1-z_2|+ \mathcal{O}(\epsilon ^2)\Big) \Big( \frac{4\pi}{\epsilon} + \mathcal{O}(\epsilon^0)\Big)
\end{equation}
and one sees the finite part of $I_0$ yielding the logarithmic piece $\sim 8\pi \log|z_1-z_2|$, as expected. 

Plugging this back in the action, one finds that the second term in (\ref{tr}) gives a contribution
\begin{equation}
e^{-\frac{\lambda_0}{\pi}|p_1|^2I_0}\sim e^{-16\lambda_0|p_1|^2\log|z_1-z_2|}=|z_1-z_2|^{-16\lambda_0|p_1|^2} ,   \label{La313macho}
\end{equation}
where the symbol $\sim $ here means that the term in $I_{\epsilon}$ that diverges when $\epsilon $ tends to zero is being omitted, as it can be absorbed in the wave function renormalization of the vertex operators (which need to be rescaled by a factor $e^{{2}(\lambda_0/{\epsilon })|p_i|^2}$). What (\ref{La313macho}) implies is that there is an anomalous dimension contribution coming from the presence of the deformation operator in the action. In other words, the operators get renormalized and the conformal dimension $\Delta_0$ of these operators receives an anomalous correction\footnote{In a paper \cite{Nuevo} that appeared today in arXiv, the authors also derive the formula (\ref{La319}) for the anomalous dimension and they discuss the spectrum of the deformed theory in relation to its non-local properties.}
\begin{equation}
\Delta_0\to \Delta=\Delta_0 +4\lambda_0|p|^2 \ \ \ \text{with}\ \ \ \Delta_0=\frac{h(1-h)}{k} . \label{La319}
\end{equation}

This is similar to what happens, for example, in the Thirring model, where the conformal dimension of the composite operator $(\bar{\psi}\psi )$ receives an anomalous correction due to the quartic fermionic interaction. Such correction actually comes from a logarithmically divergent integral like $I_0$ when correcting the 2-point function.

Then, one finds that that the presence of operator (\ref{Liouville}) in the action reduces the problem of computing (\ref{GHJ}) to that of evaluating a Liouville CFT 2-point function in presence of (\ref{Liouville}) for the appropriate value of the Liouville momenta $\alpha=b(1-h+b^{-2}/2)$ and central charge $c=1+6(b+1/b)^2$ (see \cite{HS} for the details, in particular for the explanation on how the background charge suffers a shift in $1/b$ when performing (\ref{shift})). The final result reads
\begin{equation}
\Big\langle {\Phi}_{h}(p|z_1) {\Phi}_{h}(-p|z_2)\Big\rangle  = |z_1-z_2|^{-4\Delta_0-16\lambda_0|p|^2}\frac{|p|^{4h-2}}{\pi}\Big( \frac{M_0\Gamma(\frac{1}{k})}{\Gamma(1-\frac{1}{k})}\Big)^{2h-1}\frac{\Gamma(1-2h)\Gamma(1-\frac{2h-1}{k})}{\Gamma(2h-1)\Gamma(\frac{2h-1}{k})}  \label{final}
\end{equation}
where the vertex operators $\hat{\Phi}_{h}$ have to be understood as the renormalized ones. 

It is worthwhile emphasizing that, despite we are using Wakimoto variables, no approximation has been done in deriving (\ref{final}). This is an exact path integral calculation. 

\section{Alternative derivation: perturbation theory}

The salient features in the derivation of the 2-point function (\ref{final}) are two: First, the fact that these observables can be explicitly integrated out in presence of the deformation. Second, the anomalous $p$-dependent correction to the conformal dimension. In order to further analyze the origin of this renormalization effect, let us now give an alternative derivation of the 2-point function and see how (\ref{SV0}) arises in that case. To do so, let us consider a different basis for the vertex operators. Consider \cite{BB}
\begin{equation}
\hat{\Phi}_{h,\ell ,\bar{\ell }} (z) = \gamma^{h-1-\ell } \bar{\gamma}^{h-1-\bar\ell } e^{\sqrt{{2}/{k}}(h-1)\phi}\label{OPA}
\end{equation}
which relates to the exponential basis (\ref{24}) as follows
\begin{equation}
\Phi_{h}(p|z) = \sum_{n=0}^{\infty} \sum_{m=0}^{\infty} c_n \bar{c}_m\ \hat{\Phi}_{h,h-n-1,h-m-1} \ ,  \label{316}
\end{equation}
with
\begin{equation}
c_n = \frac{(ip)^n}{\Gamma(n+1)} \ , \ \ \ \ \bar{c}_n = \frac{(i\bar{p})^n}{\Gamma(n+1)} \ . \label{317}
\end{equation}

The OPE between the current $J^-$ and the operators (\ref{OPA}) is
\begin{equation}
J^-(z)\hat{\Phi}_{h,\ell ,\bar{\ell }} (z_i) \simeq \frac{(h-1-\ell )}{(z-z_i)} \hat{\Phi}_{h,\ell +1,\bar{\ell }} (z_i) + ...
\end{equation}

After integrating over the zero modes of the fields and performing the Wick contractions, the 2-point correlation function of two operators (\ref{OPA}) is found to be
\begin{eqnarray}
\Big\langle \hat{\Phi}_{h,\ell_1 ,\bar{\ell}_1}(0) \hat{\Phi}_{h,\ell_2 ,\bar{\ell}_2}(1)\Big\rangle  &=& 
\frac{\Gamma(-s)M_0^s}{\pi^{s}}
\sum_{t=0}^{\infty }\Big[ 
 \frac{\Gamma(h-\ell_1)\Gamma(h-\ell_2)}{\Gamma(1-h+\bar{\ell}_1)\Gamma(1-h+\bar{\ell}_2)} \mathcal{I}_{s-1}(h,k) \times \nonumber \\
&& \ \ \ \ \ \ \ \times  \frac{\delta(t+\ell_1+\ell_2)}{\Gamma(t+1)} \Big(\frac{\lambda_0}{\pi} \int d^2z |z|^{-2}|z-1|^{-2} \Big)^t \Big] \label{Huna}
\end{eqnarray}
where $s=2h-1$ and 
\begin{equation}
\mathcal{I}_s(h,k) = \int_{\mathbb{C}^{s}} \prod_{r=1}^{s} d^2w_r |w_r|^{\frac{4(h-1)}{k}-2} |w_r-1|^{\frac{4(h-1)}{k}-2} \prod_{r=1}^s\prod_{t=1}^{r-1} |w_r-w_{t}|^{-\frac{4}{k}}.\label{Is}
\end{equation}

In the expression above, the factor $\Gamma(-s)$ arises through the integration over the zero-mode $\phi_0$ in the path integral, which sets the precise amount, $s=2h-1$, of screening operators $\int\beta\bar{\beta}e^{-\sqrt{k/2}\phi}$ to be inserted; see \cite{BB,GN3} for details. This integration also produces the factor $(M_0/\pi)^s$. Notice that the expression above makes sense in principle only for $2h\in\mathbb{Z}_{> 1}$, for which the amount of integrals to be performed in (\ref{Is}) turns out to be a positive integer. However, the resulting expression admits a well-known analytic extension to other values of $s$ and the result is valid for general values of $h$. In the expression above, we have also invoked $PSL(2,\mathbb{C})$ invariance to fix the vertex operators at $z_1=0$ and $z_2=1$. In order to fully stabilize the projective symmetry and cancel out the volume of the Killing conformal group in the path integral, we have also fixed the $s^{\text{th}}$ interaction operator $w_s=\infty $; this is why $\mathcal{I}_{s-1}(h,k)$ appearing in (\ref{Huna}) turns out to be an integral over $s-1$ and not over $s$ variables. The sum over the index $t$ stands for different values of $\ell_1+\ell_2$. Because of the presence of the deformation operator (\ref{D}), which does not commute with the $U(1)$ Cartan element of $SL(2,\mathbb{R})$, the quantity $t=-\ell_1-\ell_2$ is not necessarily conserved and one has to sum over its possible values; this explains the sum over $t$ as well as the factor $(\lambda_0/\pi)^t/t!$. The factorial follows from the different ways of ordering the insertions of $D^t$ operator when treating (\ref{D}) as a perturbation.  

Remarkably, integral (\ref{Is}) can be explicitly solved to produce a relatively simple formula in terms of $\Gamma$-functions \cite{DF2}. Integral
\begin{equation}
\mathcal{I}_1(h=1)=\int_{\mathbb{C}} d^2z\ |z|^{-2}|z-1|^{-2},
\end{equation}
is, in contrast, divergent. In fact, this is exactly the integral we have found before, in (\ref{SV0}), which here appears in all terms of the series conveniently accompanied with the same power of $\lambda_0$. In this basis, integral $I_0=\mathcal{I}_1(h=1)$ has its origin in the contraction of the $\beta $ fields of operators $D$ with the $\gamma$ fields of the vertex operators $\Phi_{h,\ell,\bar\ell}$. More precisely, in the limit $h=1+k\epsilon /2\to 1$ ($\epsilon \to 0$) the integral $\mathcal{I}_1(h,k)$ goes like $\sim 4\pi /\epsilon $. Therefore, we find exactly the same renormalization as before, with the logarithmic divergence leading to the anomalous dimension (\ref{La319}). After integrating over $w_r$ and considering the renormalization of $\lambda_0$, we find
\begin{eqnarray}
\Big\langle \hat{\Phi}_{h,\ell_1 ,{\ell}_1}(0) \hat{\Phi}_{h,\ell_2 ,{\ell}_2}(1)\Big\rangle &=& 
\frac{1}{\pi }\Big( M_0 \frac{\Gamma(\frac{1}{k})}{\Gamma(1-\frac{1}{k})}\Big)^{2h-1}\frac{\Gamma(1-2h)\Gamma(1-\frac{2h-1}{k})}{\Gamma(2h-1)\Gamma(\frac{2h-1}{k})}\times \nonumber \\
&& \ \  \times \sum_{t=0}^{\infty }\frac{\lambda_0^tI_0^t}{\pi^t t!}
\frac{\delta(t+\ell_1+\ell_2)\Gamma(h-\ell_1)\Gamma(h-\ell_2)}{\Gamma(1-h+\ell_1)\Gamma(1-h+\ell_2)} , \label{Cuco}
\end{eqnarray}
which is the result for the 2-point function in the basis (\ref{OPA}). Notice that the first step in the sum ($t=0$) correctly reproduces the 2-point function of the undeformed $SL(2,\mathbb{R})$ WZW theory, which corresponds to $\lambda_0 =0$. In other words, correlator (\ref{Cuco}) generalizes the result for the reflection coefficient of strings in AdS$_3$. On the other hand, the 2-point function (\ref{final}), which corresponds to operators of the basis (\ref{24}), in the case $\lambda_0=0$ can be relevant to describe string interactions on the massless Ba\~{n}ados-Teitelboim-Zanelli geometry \cite{Ross}. Using (\ref{316})-(\ref{317}) and properties of the $\Gamma$-functions, such as $\Gamma(n-z)\Gamma(1+z-n)=(-1)^{n}\Gamma(1+z)\Gamma(-z)$, one can easily recover (\ref{final}) from (\ref{Cuco}). In fact, it is easy to verify that, due to the presence of the function $\delta(t+\ell_1+\ell_2) $ in each term of the sum over $t$, (\ref{Cuco}) before renormalization reproduces the behavior
\begin{eqnarray}
\Big\langle {\Phi}_h(p|0) {\Phi}_h(-p|1)\Big\rangle \sim \  e^{-\frac{\lambda_0I_0|p|^2}{\pi }}
\frac{|p|^{4h-2}}{\pi }\Big( M_0 \frac{\Gamma(\frac{1}{k})}{\Gamma(1-\frac{1}{k})}\Big)^{2h-1}\frac{\Gamma(1-2h)\Gamma(1-\frac{2h-1}{k})}{\Gamma(2h-1)\Gamma(\frac{2h-1}{k})}, 
\end{eqnarray}
consistently with (\ref{final}). Anomalous dimension follows from (\ref{La313macho}) after reintroducing the dependence on $z_{1,2}$. 

\section{Pole structure}

Then, we have computed the 2-point function in the presence of the deformation parameter for operators in two different basis and by means of two different methods. The results are (\ref{final}) and (\ref{Cuco}). In both basis, the 2-point function exhibits singularities at 
\begin{equation}
h = \frac{n}{2} \ , \ \ \ \ n\in \mathbb{Z}_{>0} \label{primeros}
\end{equation}
and at
\begin{equation}
h =\frac{1}{2}+ \frac{n}{2} k \ , \ \ \ \ n\in \mathbb{Z}_{>0} . \label{segundos}
\end{equation}

On general grounds, one expects the $N$-point correlation function to develop singularities at $\sum_{i=1}^N h_i=n+N-1  $ and at $\sum_{i=1}^N h_i=nk+N-1  $, with $ n\in \mathbb{Z}_{> 0}$, along with other infinite poles. These pole conditions are of course also present in the undeformed case, $\lambda_0=0$. However, one has to be reminded of the fact that the index $h$, as a function of the dimension $\Delta_0 $, also depends on the quantity $\lambda_0|p|^2$ through equation (\ref{La319}). In other words, the dependence on $h$ in the analytic expression of the 2-point function (\ref{final}) is ultimately $p$-dependent, in virtue of the mass-shell condition (see (\ref{POI1})-(\ref{POI2}) below). 

The poles (\ref{primeros}) have their origin in the factor $\Gamma(-s)$ in (\ref{Huna}), which diverges at $s\in\mathbb{Z}_{\geq 0}$. For kinematic configurations such that $s=\sum_{i=1}^{N}h_i+1-N>0$, the $N$-point correlators are dominated by the region in which the exponential self-interaction (the Liouville-type wall) is negligible. This is because that is the region where the integration over the zero-mode $\phi_0$ principally contributes. On the contrary, when $s=\sum_{i=1}^{N}h_i+1-N<0$, the correlators receive the principal contributions from the vicinities of the the wall. This is similar to what happens in other 2-dimensional string theory examples \cite{DiFK}, and the interpretation has to be similar too: From the bulk geometry point of view, poles (\ref{primeros}) have to be understood as resonances in the scattering of states with the microscopic constituents of the wall.  

The interpretation of poles (\ref{segundos}) is more subtle. As pointed out in \cite{StringyHorizons}, the factor responsible for such poles exhibits at high energy a peculiar functional form, a functional form that differs from the semi-classical result for a reflection coefficient along the semi-infinite direction $\phi$. Indeed, using Stirling approximation one finds the following result in the large $p_{\phi}/\sqrt{k}$ limit 
\begin{equation}
\frac{\Gamma(1-\frac{2h-1}{k})}{\Gamma(\frac{2h-1}{k})} \sim ip_{\phi }\ e^{-\frac{2ip_{\phi }}{\sqrt{k}}(\log p_{\phi }-1-\frac{1}{2}\log k)}  ,\label{PhaseShift}
\end{equation}
recall $2h-1= i \sqrt{k}p_{\phi }$.
In the case in which the gauged $SL(2,\mathbb{R})/U(1)$ WZW model is considered to describe 2-dimensional string theory on the Euclidean black hole geometry, the prefactor (\ref{PhaseShift}) in the 2-point function was interpreted as an anomalous phase-shift that captures finite $\alpha '$ effects in the near horizon geometry \cite{StringyHorizons}. We see here this factor appearing as well, as it is characteristic of the $SL(2,\mathbb{R})$ component of the CFT$_2$. Then, one could feel tempted to give to it an interpretation in terms of how the high-momentum states experience the $\phi\to-\infty$ zone of the linear dilaton direction in the deformed theory. However, whatever interpretation for the large momentum behavior of this factor is to be proposed, this should also be applicable to the undeformed case $\lambda_0 =0$.

There is also interesting physics associated to the factor $\sim |p|^{4h-2}$ to which the 2-point function of operators (\ref{24}) results proportional. This is discussed in detail in \cite{Nuevo}.

\section{Relation to the coset construction and spectrum}

Now, let us discuss the spectrum of the theory. Following \cite{GIK2}, one can describe the bulk theory proposed in \cite{GIK1} as a coset, starting with the background
\begin{equation}
\mathbb{R}\times S^1 \times \text{AdS}_3 \times S^3 \times T^4 \label{bac}
\end{equation}
and gauging the current $J^-$ of $SL(2,\mathbb{R})$. The bulk geometry obtained through this gauging procedure is the geometry of strings and five-branes discussed in \cite{Pelc}. To implement the gauging one introduces two free fields, $T$ and $X$, representing the $\mathbb{R}$ and the $S^1$ directions of (\ref{bac}), respectively; the first of these fields, $T$, represents a timelike direction, namely $\langle X(z) X(0) \rangle = - \langle T(z) T(0) \rangle = -2\log |z|$. 

After dressing operators (\ref{24}) with the $T$, $X$ auxiliary fields that are employed to realize the coset, and including the internal space, one gets 
\begin{equation}
\hat{\mathcal{O}}_h(p)= \int d^2z \ e^{-\varphi -\bar{\varphi}} \Phi_h(p|z) e^{-i\sqrt{\frac{2}{k}}(wT(z)+p_x X(z))} \mathcal{O}(z)
\end{equation}
where $\varphi $, $\bar{\varphi}$ are the superconformal ghosts, which we will ignore hereafter. $\mathcal{O}$ corresponds to the primary operator of dimension $\Delta_{\mathcal{O}}$ in the CFT$_2$ defined on the $S^3 \times T^4$ sub-manifold. Notice that here we are focusing only on the bosonic components of the AdS$_3$ piece of the geometry, although the fact that we are considering the supersymmetric theory is reflected, for example, in the fact that $k$ does not suffer any shifting in the expressions for the vertex operators and their conformal dimensions.  

The coset is realized by gauging the null current, which we parameterize by the improved current
\begin{equation}
\hat{J}^- (z)= \sqrt{\frac k2} (\partial X  - \partial T )+i \varepsilon \partial v \ e^{iv-u} 
\end{equation}
where $\varepsilon$ is related to the crossover scale $\lambda_0$. In fact, one can verify easily that the following OPE holds
\begin{equation}
\hat{J}^-(z)\Phi_h(p|z_i) e^{-i\sqrt{\frac{2}{k}}(wT(z_i)+p_x X(z_i))} \simeq \frac{i}{(z-z_i)} (w+p_x-\varepsilon p) \Phi_h(p|z_i) e^{-i\sqrt{\frac{2}{k}}(wT(z_i)+p_x X(z_i))} + ... \nonumber
\end{equation}
which becomes regular provided the constraint $w+p_x-\varepsilon p=0$ is obeyed; analogously, $w-{p}_x-\varepsilon \bar{p}=0$. 

Similarly, the stress-tensor ${\mathcal T}$ is defined by adding to the one derived from (\ref{S}) the $T$, $X$ contribution $+\frac 12 (\partial T )^2 -\frac 12 (\partial X )^2$. Then, the following OPE is easily be obtained
\begin{eqnarray}
{\mathcal T}(z)\ \Phi_h(p|z_i)e^{-i\sqrt{\frac{2}{k}}(wT(z_i)+p_x X(z_i))} &\simeq & \frac{\Delta }{(z-z_i)^2}  \Phi_h(p|z_i)e^{-i\sqrt{\frac{2}{k}}(wT(z_i)+p_x X(z_i))} + \nonumber \\
&+& \frac{1}{(z-z_i)} \partial\ \Phi_h(p|z_i)e^{-i\sqrt{\frac{2}{k}}(wT(z_i)+p_x X(z_i))}  + ... \nonumber
\end{eqnarray}
with
\begin{equation}
\Delta = -\frac{h(h-1)}{k} + \frac{\alpha '}{4} (p^2_x - w^2) = -\frac{h(h-1)}{k} + 4\lambda_0|p|^2 , \label{44}
\end{equation}
for the appropriate identification $\varepsilon = 2\sqrt{k\lambda_0}$ \cite{GIK2}. This reproduces (\ref{La319}). 

Once (\ref{44}) has been obtained, the study of the spectrum follows from the analysis of \cite{GIK2}. To describe the spectrum on the $\mathbb{R}\times S^1$ sup-space parameterized by $T$, $X$, one splits the left- and the right-moving components as
\begin{equation}
e^{i\sqrt{\frac 2k}p_x X} = e^{i\sqrt{\frac 2k}p_L X_L+ i\sqrt{\frac 2k}p_R X_R} 
\end{equation}
where $k=R^2/\alpha '$, having chosen here $R=2$ for convention. This permits to describe the Kaluza-Klein and the winding modes along the compact direction; namely
\begin{eqnarray}
p_L&=& \frac{n}{2}+\frac{k\omega }{2} = \frac{n}{R} + \frac{R\omega }{\alpha '} \nonumber \\ 
p_R&=& \frac{n}{2}-\frac{k\omega }{2} = \frac{n}{R} - \frac{R\omega }{\alpha '} .
\end{eqnarray}

The string theory spectrum then follows from the Virasoro constraint of the supersymmetric theory, $\Delta+\bar{\Delta}=1$; namely
\begin{eqnarray}
\Delta &=&-\frac{h(h-1)}{k}+\frac{\alpha ' }{4} (p_L^2 - w ^2)+\Delta_{\mathcal{O}}=\frac{1}{2}  \label{POI1} \\
\bar{\Delta }&=&-\frac{h(h-1)}{k}+\frac{\alpha ' }{4} (p_R^2 - w ^2)+\bar{\Delta}_{\mathcal{O}}=\frac{1}{2} .\label{POI2} 
\end{eqnarray}

This mass-shell condition yields the energy spectrum
\begin{equation}
w=\sqrt{\Big( \frac nR\Big)^2 + \Big( \frac {\omega R}{\alpha ' }\Big)^2 + 
\frac{1}{R^2}+\frac{p_{\phi}^2}{\alpha '}+\frac{2}{\alpha '} 
\Big(\Delta_{\mathcal{O}} + \bar{\Delta}_{\mathcal{O}}-1\Big) }\label{msc}
\end{equation}
together with the level matching condition for the internal spin $\Delta_{\mathcal{O}} - \bar{\Delta}_{\mathcal{O}} = -n\omega .$

Formula (\ref{msc}) yields the energy spectrum studied in \cite{GIK2}. This, after choosing the correct ground states and elaborating a little, leads to an interesting discussion about the Hagedorn spectrum and the thermodynamics of the deformed theory. We refer to \cite{GIK2} for the details. 

Going back to the formula (\ref{La319}) for the anomalous dimension, and considering this in relation to the mass-shell condition (\ref{POI1})-(\ref{POI2}), one finds
\begin{equation}
h=\frac{1}{2}+\frac{1}{2}\sqrt{1+4k\Big( \Delta_{\mathcal O}+\frac{1}{2}\Big) +16k\lambda_0 p^2} ,
\end{equation}
which manifestly shows the dependence on $p$. The solution with $h'=1-h\geq 0$ is also possible within the window
\begin{equation}
0\geq 4\lambda_0p^2+\Delta_{\mathcal O}+\frac{1}{2}\geq -\frac{1}{4k}.
\end{equation}

In order to guarantee that $h\in \mathbb{R}$, one needs to impose the condition
\begin{equation}
4\lambda_0p^2+\Delta_{\mathcal O}+\frac{1}{2}\geq -\frac{1}{4k};
\end{equation}
which corresponds to the values of $h$ that, in the undeformed theory, belong to discrete representations of the universal covering of $SL(2,\mathbb{R})$. On the other hand, for the range
\begin{equation}
4\lambda_0p^2+\Delta_{\mathcal O}+\frac{1}{2}< -\frac{1}{4k}.
\end{equation}
the index $h$ takes values corresponding to the continuous series (\ref{LAGHY}).

\section{Higher-point correlation functions}

To conclude, let us make a few remarks. The first is about the 3-point correlation function, which can also be integrated explicitly. This corresponds to the deformation by operator (\ref{D}) of the $SL(2,\mathbb{R})$ WZW 3-point function in the exponential basis (\ref{24}). This can be done by adapting the results of \cite{HS} to the deformed theory. The presence of the deformation operator in the action actually produces a divergent piece
\begin{equation}
I_{0 }\sim  \int d^2z\ |z-y|^{2}\prod_{i=1}^3|z-z_i|^{-2}, \label{ConfInt}
\end{equation}
with $p_1+p_2+p_3=0$. This is similar to the one we found for the 2-point function, except for the fact that here there is an additional insertion at 
\begin{equation}
y=-\frac{p_1z_2z_3+p_2z_3z_1+p_3z_1z_2}{p_1z_1+p_2z_2+p_3z_3}.
\end{equation} 

Integral (\ref{ConfInt}) appears because, when a third vertex is inserted in the correlator, the solution for field $\beta$ gets modifies as follows
\begin{equation}
\beta(z)=i(z-y)\sum_{i=1}^3p_iz_i\ \prod_{j=1}^3(z-z_j)^{-1},
\end{equation}
and, when evaluated on the term $D$ in the action, it yields (\ref{ConfInt}). Conformal integral (\ref{ConfInt}) can be solved by resorting to the following functional identity
\begin{eqnarray}
&&\int_{\mathbb{C}}d^2z \ |z-y|^{-4-2\sum_{i=1}^3a_i}\prod_{i=1}^3|z-z_i|^{2a_i} = \frac{\Gamma(-1-\sum_{i=1}^{3}a_i)}{\Gamma(2+\sum_{i=1}^{3}a_i)}\prod_{i=1}^3\frac{\Gamma(1+a_i)}{\Gamma(-a_i)} \Big\vert \frac{z_1-z_2}{z_3-y}\Big\vert ^{2+2a_1+2a_2} \times \nonumber \\
&&\ \ \ \ \ \ \ \ \ \ \ \ \ \ \ \ \times \int_{\mathbb{C}} d^2z |z-y|^{-2-2a_3} |z-z_1|^{-2-2a_2}|z-z_2|^{-2-2a_1}|z-z_3|^{2+2\sum_{i=1}^{3}a_i}
\end{eqnarray}
considering the particular case $a_1=-1+\epsilon$, $a_2=-1-\epsilon$, $a_3=-1+2\epsilon$. Following the steps above, this yields a closed expression. The result can be conveniently expressed in terms of a 4-point function in Liouville theory. 

$N$-point correlation functions with $N\geq 3$ can also be expressed in terms of Liouville theory correlators multiplied by the exponential of a conformal integral like (\ref{ConfInt}) in which more products $\prod_{i=1}^{N-2}|z-y_i|^2\prod_{j=1}^{N}|z-z_j|^{-2}$ appear in the integrand, with the complex variables $y_i,\bar{y}_i$ being functions of $z_j,\bar{z}_j$ and $p_j,\bar{p}_j$. This follows as a corollary of Riemann-Roch theorem. More precisely, it follows from the fact that $\beta $ and $\bar{\beta }$ are meromorphic 1-differential on a genus-zero surface, so their number of zeroes (of multiplicity one) exceed in 2 their number of (single) poles, producing such integrand. This means that for $N$-point function with arbitrary $N$ the operator $D$ will produce a divergence as the one that accounts for (\ref{La319}) in the 2-point function.   

\section{Correlators redux: conjugate representation}

It seems convenient to think of the deformation (\ref{D}) in relation to the conjugate representations that the WZW theory admits when described as a product 
\begin{equation}
\frac{SL(2,\mathbb{R})}{U(1)} \times U(1) ,\label{product}
\end{equation}
as proposed in \cite{GN3}. This is also related to the duality studied in \cite{StringyII}. In fact, according to the representation of \cite{GN3} the operator (\ref{SD}) can be associated to the operator 
\begin{equation} 
2{\lambda_0 } \int d^2z\ e^{i\sqrt{\frac 2k}X+i\sqrt{\frac 2k}T} \label{esoy}
\end{equation} 
on the product space (\ref{product}), where $X$ is a space-like free boson that realizes the $U(1)$ that is being modded out and $T$ is a time-like free boson that realizes the extra $U(1)$ factor. Field $X$ presents a background charge $(Q_X/\pi)\int d^2z\mathcal{R}X$ with $Q_X^2=k/2$, so that the operator (\ref{esoy}) has dimension 1. In this conjugate representation, the vertex operators $\hat{\Phi}_{h_i,\ell_i ,{\ell}_i}$ result proportional to $ e^{i\sqrt{2/k}(\ell_i +k/2)X+i\sqrt{2/k}\ell_i T}$, yielding a contribution ${\lambda_0 } I_0/{\pi}$ with 
\begin{equation} 
 I_0=  \int d^2z\ \prod_{i=1}^N|z-z_i|^{-2}\label{GHH}
\end{equation} 
when contracted with operator (\ref{esoy}), as in (\ref{Huna}). This type of representation can provide a simpler way of dealing with the deformed theory. To see how this works explicitly, one can consider again the 2-point function and write the operators $\hat{\Phi}_{h,\ell,\ell}$ in the conjugate representation \cite{GN3}
\begin{equation}
\hat{\Phi}_{h,\ell,{\ell}}=(\beta \bar{\beta} )^{h+\ell-1/2}e^{-\sqrt{\frac{2}{k}}(1-h+k/2)\phi}e^{i\sqrt{\frac{2}{k}}(\ell+k/2)X+i\sqrt{\frac{2}{k}}\ell T},
\end{equation}
which has the advantage of not involving the $\gamma$ field. As a consequence, the ghosts fields decouple and the 2-point function $\langle \hat{\Phi}_{h,\ell_1,{\ell}_1}(z_1) \hat{\Phi}_{h,\ell_2,{\ell}_2}(z_2)\rangle $ reduces to that of two exponential operators $e^{\sqrt{2}\alpha\phi}$ with momenta $\alpha = (h-1-k/2)/\sqrt{k}$ in Liouville field theory at central charge $c=1+6(b+1/b)^2$, with $b=-1/\sqrt{k}$, all multiplied by a factor $\sum_{t=0}^{\infty }\delta(\ell_1+\ell_2+t)(\lambda_0 I_0/\pi)^t/t!$ coming from (\ref{GHH}). The compensation condition $\ell_1+\ell_2+t=0$ is realized because of the presence of the background charge $Q_X$. Again, the exponentiation of $I_0$ yields the logarithmic divergence responsible for the anomalous correction to the conformal dimension. The final result exactly reproduces (\ref{final}).

\section{Final remarks}

As shown here, for the model deformed by operator (\ref{D}) the correlation functions can be explicitly computed and this leads to the determination of the spectrum in agreement with \cite{GIK2}. However, open questions still remain. For instance, it would be interesting to fully understand the connection between the deformation (\ref{D}) proposed in \cite{GIK1} and the renormalization group flow triggered by the standard $T\bar{T}$-deformation recently considered in references \cite{SmirnovZamolodchikov, Cavaglia}. 

One might also ask whether other solvable variations of the model are possible. In fact, one can also imagine other deformations of the theory (\ref{SD}) that are integrable as well, in the sense that their correlation functions can be in principle computed explicitly. For instance, one could consider the family of theories introduced in \cite{Ribault}, which are deformations of the $SL(2,\mathbb{R})$ WZW that consist in replacing the fifth term in (\ref{SD}) by the operator $\lambda_0\int d^2z (-2 \beta\bar{\beta })^{1+\eta } e^{-\sqrt{{2}/{k}}\phi } $ along with a change in the background charge in the second term, namely $\sqrt{2/k}\int d^2z\mathcal{R}\phi\to (\sqrt{2/k}-\sqrt{2k}\eta)\int d^2z\mathcal{R}\phi$, with $\eta \in\mathbb{R}$. After integrating the $\beta $, $\bar{\beta}$ fields, this yields higher-derivative interactions provided $\eta \neq 0$. These theories also have operator (\ref{D}) as a worldsheet marginal deformation and the correlation functions can also be written explicitly using the techniques discussed here. However, the geometric interpretation of these theories as string $\sigma$-models is unclear precisely because of the higher-derivative terms. 

Another interesting problem would be studying the analytic properties of the 2-point function (\ref{final}) for operators (\ref{24}) and trying to infer from this some aspects of the theory related to its non-locality. Interesting results on this have been obtained in the paper \cite{Nuevo}, which appeared in arXiv today.

It would also be interesting to understand the theory for negative values of $\lambda_0$ and answer the question whether a vacuum exists in that case. The relation to other holographic realization of $T\bar{T}$-deformed theories, such as the one of \cite{Verlinda, Shyam}, is also an interesting problem to investigate.

\[
\]

The author is grateful to David Kutasov for discussions. The work of the author is supported by the NSF through grant PHY-1214302.

\end{document}